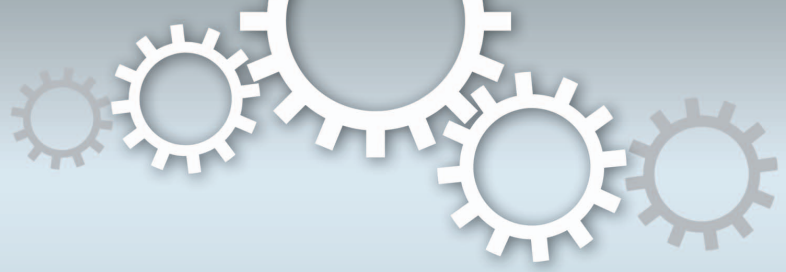

OPEN

# Itinerant magnetism in doped semiconducting $\beta$-FeSi$_2$ and CrSi$_2$

David J. Singh & David Parker

Materials Science and Technology Division, Oak Ridge National Laboratory, Oak Ridge, TN 37831-6056, USA.

Novel or unusual magnetism is a subject of considerable interest, particularly in metals and degenerate semiconductors. In such materials the interplay of magnetism, transport and other Fermi liquid properties can lead to fascinating physical behavior. One example is in magnetic semiconductors, where spin polarized currents may be controlled and used. We report density functional calculations predicting magnetism in doped semiconducting $\beta$-FeSi$_2$ and CrSi$_2$ at relatively low doping levels particularly for $n$-type. In this case, there is a rapid cross-over to a half-metallic state as a function of doping level. The results are discussed in relation to the electronic structure and other properties of these compounds.

$\beta$-FeSi$_2$ is a semiconductor that can be grown in high quality form on bulk Si and is of interest for electro-optic applications. Importantly, it can be doped both $p$-type and $n$-type[1–7]. CrSi$_2$ is also a semiconductor that can have both carrier signs[8,9], and can be grown on different faces of Si[10]. In contrast to $\beta$-FeSi$_2$, CrSi$_2$ has a non-centrosymmetric space group.

There have been a number of prior band structure calculations for both $\beta$-FeSi$_2$ and CrSi$_2$[11–15], both of which are of potential interest as thermoelectric materials[16–19]. These studies show a wide set of Si $s$ and $p$ bands on top of which there are much narrower transition metal $d$ bands. Both compounds are semiconducting due to the details of the hybridization between the Si and transition metal states.

The result is that these are semiconductors with heavy mass, transition metal derived bands at the band edges. This implies that, relative to standard light mass semiconductors, the density of states will be high when doped – a fact that is important for thermoelectric performance[17–19]. Additionally, when these compounds are doped the resulting electron or hole pockets involved in conduction have substantial anisotropy. This is of importance because the direction averagings for the density of states and thermopower are different from that in the direction dependent conductivity, leading to the possibility of doping and good conduction in a material that from a density of states or thermopower perspective has very heavy bands[18–21].

We note that depending on the doping level and the details of the amount of transition metal character in the band-edge states, this can potentially lead to satisfaction of the Stoner criterion for ferromagnetism and a ferromagnetic ground state. In fact, FeGa$_3$, which has a qualitatively similar band character, does become a ferromagnet when heavily doped $n$-type by alloying with Ge[22,23], consistent with this scenario[24]. Interestingly, there have been reports that suggest magnetism or nearness to magnetism in FeSi$_2$ based nanostructures[25,26]. The purpose of this work is to examine this possibility for FeSi$_2$ and CrSi$_2$.

## Results

As mentioned, the possibility of magnetism will depend on the details of the electronic structure near the band edges. The densities of states, $N(E)$ and transition metal $d$ projections in the band edge region are shown in Fig. 1(a). Fig. 1(b) shows the corresponding carrier density as a function of energy for the two compounds, while Figs. 2(a) and 2(b) show the projections onto the various $d$ orbitals of the transition metals.

Both compounds show high values of the density of states close to the band edges, particularly for $\beta$-FeSi$_2$ and for the conduction bands. This parallels the expected thermoelectric performance of these two compounds, which is most favorable for $n$-type. In particular, values of $N(E)$ well in excess of 2.5 eV$^{-1}$ per formula unit are seen, which is the value at which a Stoner instability may be expected if the $d$ character is high, as is the case here (note that the Stoner instability depends on the value of $N(E_F)I$, with $N(E_F)$ converted to per spin, per atom basis and $I$ an interaction parameter that is roughly 1 eV for 3$d$ transition elements and is much lower for Si; thus the Si contribution component of the density of states does not contribute significantly to this instability)[27]. More specifically, for $\beta$-FeSi$_2$, $N(E)$ rises sharply immediately above the conduction band minimum (CBM). In contrast, there is a lighter band with Si character at the valence band maximum (VBM) as has been discussed in previous work[13,15,17,19]. This leads to a low $N(E)$ im-mediately below the band edge, followed by a heavier behavior starting at ~0.1 eV binding energy and finally a sharp increase, comparable to that at the CBM, starting at ~0.2 eV binding energy. CrSi$_2$ shows a sharply increasing $N(E)$ above the CBM, though this is preceded by a shoulder, which extends for the first ~0.1 eV above the CBM. The VBM of CrSi$_2$ also shows a shoulder, and unlike the CBM does not show as sharp an increase beyond the shoulder. The projections of the density of states show that this high density of states comes from different $d$ orbitals for the conduction and valence bands, which is a characteristic of these silicides[18,21].

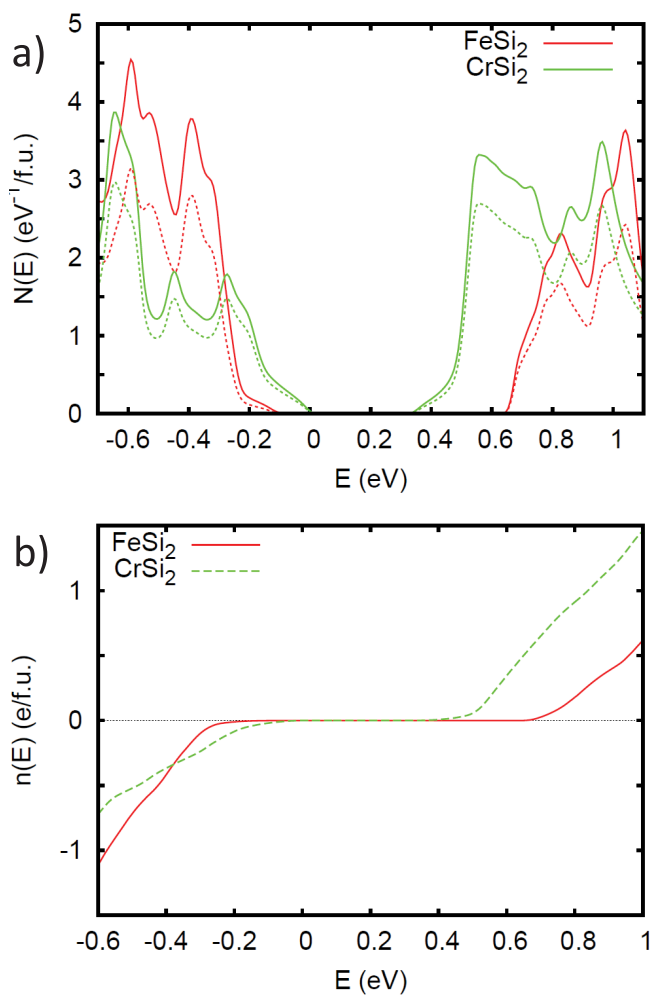

**Figure 1** | (a) Electronic density of states of non-magnetic undoped $\beta$-FeSi$_2$ and CrSi$_2$ on a per formula unit basis (solid lines) along with the transition metal $d$ projections onto the LAPW sphere (dotted lines). The energy zero is at the valence band maximum (VBM). Note that VBM of $\beta$-FeSi$_2$ consists of a light Si $p$ derived band. This gives a very low DOS, not clearly visible on this scale, near the band edge, which is at 0 eV. (b) Electron count as a function of energy on a per formula unit basis.

The main result of the present study is that this high density of states leads to Stoner type magnetism. Fig. 3 shows the calculated magnetization as a function of doping level for the two compounds. Energetics for selected doping levels are given in Table 1. Both CrSi$_2$ and $\beta$-FeSi$_2$ show an onset of ferromagnetism, followed by a rapid cross-over to a maximally spin polarized, half-metallic state for $n$-type. This cross-over can be understood within extended Stoner theory[28,29], as was discussed by Kulatov and Mazin[30]. A doping level 0.1 e/f.u. corresponds to $2.658 \times 10^{21}$ cm$^{-3}$ for $\beta$-FeSi$_2$, and $2.775 \times 10^{21}$ cm$^{-3}$ for CrSi$_2$.

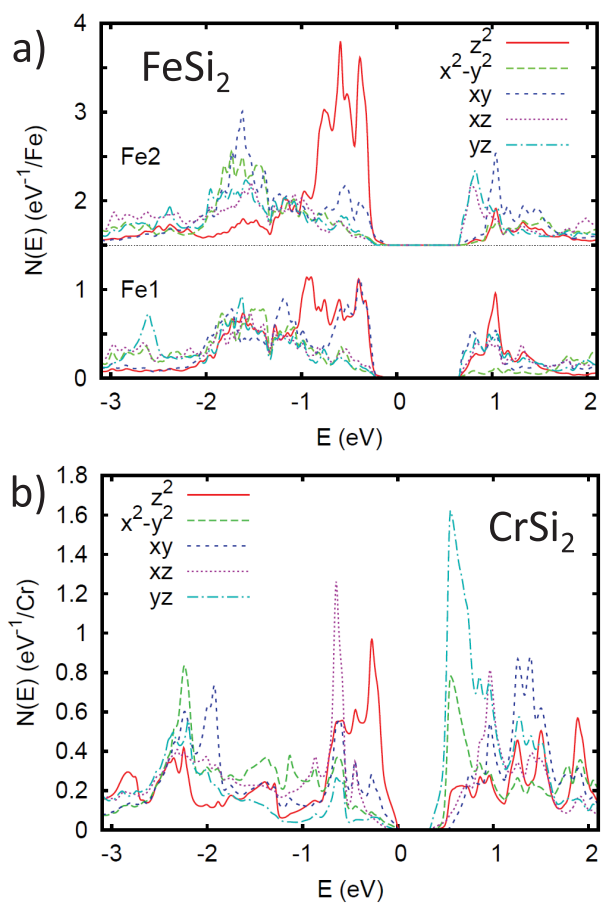

**Figure 2** | (a) Projections of the transition metal density of states onto the different $d$ orbitals for FeSi$_2$ and (b) for CrSi$_2$. Note that in these compounds the transition elements occur on sites of low symmetry and as a consequence the five $d$ orbitals are independent from the point of view of crystal field. This is in contrast to e.g. a cubic site where they would occur in two manifolds, $e_g$ and $t_{2g}$. The projections are onto the LAPW spheres and are given on a per atom basis. For FeSi$_2$ the two Fe sites are shown separately, with the Fe2 offset by 1.5 eV$^{-1}$.

Half metals are ferromagnets where one spin channel is semiconducting, while the other is metallic[31,32]. In such materials, spin flip scattering is blocked and ferromagnetic domain walls can have high resistance, leading to unusual magnetotransport phenomena such as large negative magnetoresistance. Such materials are also of interest as "spin-tronic" materials, since their electrical transport is entirely in one spin channel, and also because the magnetic excitation spectrum is different than in ordinary metallic ferromagnets, in particular as regards the Stoner continuum.

$\beta$-FeSi$_2$ and CrSi$_2$ both show weaker magnetism for $p$-type doping relative to $n$-type. In the case of CrSi$_2$, there is only a narrow composition range where magnetism is expected, $p = 0.1–0.3$ holes per formula unit. The vanishing of magnetism above this doping level was confirmed by fixed spin moment calculations (Fig. 4), which show that the energy increases monotonously as a function of magnetization for this and higher doping levels. This is a consequence of the fact that $N(E)$ shows lower values for $p$-type than for $n$-type. For FeSi$_2$, the onset of magnetic behavior is at higher doping levels than for $n$-type, and a fully spin polarized, half-metallic state is not reached. Thus the case that reaches ferromagnetism at the lowest doping is $n$-type $\beta$-FeSi$_2$.

We did additional calculations for $\beta$-FeSi$_2$ with chemical doping via replacement of one Si atom in the 24 atom (eight formula unit) cell by B, P or As. For $n$-type, this doping level of 0.125 e/f.u. (1 e/

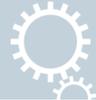

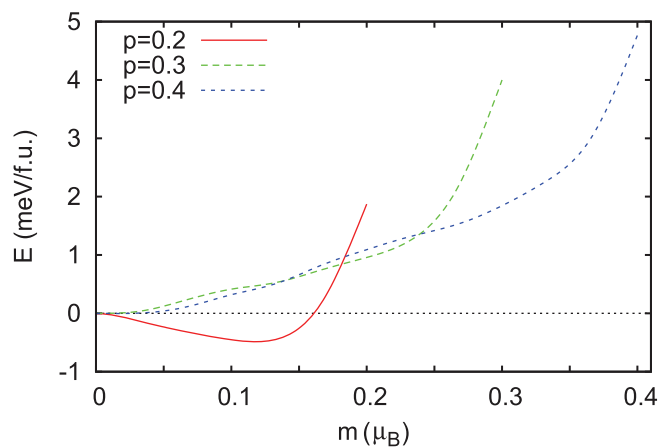

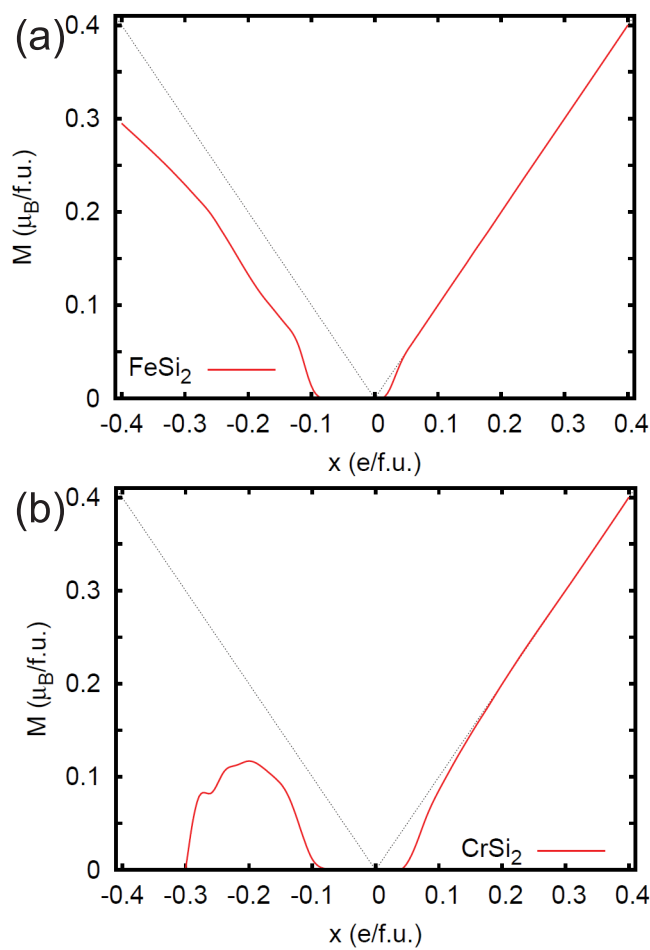

**Figure 3** | (color online) Calculated magnetization per formula unit as a function of doping level, x, for p-type (negative x, $p = -x$) and n-type (positive x, $n = x$) $\beta$-FeSi$_2$ (a, top) and CrSi$_2$ (b, bottom). The dotted lines indicate full half metallic polarization.

cell) is very close to the point where the VCA calculations predict that the magnetic state crosses over from partially polarized to half-metallic. For P and As doping we find a half-metallic state. For p-type this doping level is in the middle of the cross-over from non-magnetic to half-metallic. The VCA spin magnetization is 0.49 $\mu_B$/cell for this doping level. Our B doped cell had a calculated magnetization 0.54 $\mu_B$/cell again in close accord with the VCA.

**Figure 4** | (color online) Fixed spin moment energy as a function of magnetization for CrSi$_2$ on a per formula unit basis for various p-type doping levels.

We also checked for antiferromagnetic solutions with the VCA, but did not find any. Specifically, we did calculations seeking a state in which the nearest neighbor Fe were antiferromagnetic (these are the Fe on two different crystallographic sublattices, so strictly speaking this should be called a ferrimagnetic state although it has equal numbers of spin up and spin down Fe atoms). These calculations were done for n-type doping levels of 0.1, 0.2 and 0.4 e/f.u. In none of these cases were we able to stabilize such an antiferromagnetic state. This shows that the ferromagnetism is itinerant.

## Discussion

The itinerant mechanism for magnetism, coming from a Stoner instability of a dilute system, is different from traditional dilute magnetic semiconductors, such as Mn doped GaAs[33]. In those materials local moments exist on impurity atoms or in some cases at other point defects. Those moments are then coupled through the host lattice, typically mediated by conduction electrons. In that case, one may describe the interacton between the moments using Heisenberg or Ising models (depending on the spin and site anisotropy)[34]. Thus there are two ingredients: local moments on impurities, whose existence, properties and interaction with the host depend on the specific identity of the defect, and the host mediated interaction, which depends on details of the host, such as its identity and its doping level. Here, we find a different kind of ferromagnetism, i.e. itinerant ferromagnetism. This comes from a Stoner instability of the electron gas in the host when doped. Therefore, it does not depend on the introduction of moments on magnetic impurities, but does depend on the carrier concentration. Thus provided that one can dope carriers into these materials, the prediction does not depend on the specific dopant. For example, we find magnetism when doping with As, which is not an element that ordinarily carries a local moment in solids. We emphasize that the magnetism cannot be described by Heisenberg type models, as for example, contrary to the basis of these local moment models, the moment vanishes for imposed antiferromagnetic orderings.

As mentioned, the ferromagnetism found here in doped $\beta$-FeSi$_2$ and CrSi$_2$ is not associated with local moments on dopant atoms. It is, however, associated with a polarization of the electronic states in the doped system. As seen in the density of states, these electronic states, which are the near band edge electronic states, have dominant transition metal d chararacter. Polarization of transition metal d electrons can lead to either local moment magnetism, as in e.g. manganites, or to itinerant magnetism as in Ni$_3$Al[35]. The distinction is the extent to which moments exist independent of the specific magnetic order. In other words, in a local moment system, the similar sized moments are present for different ordering patterns, including the disordered configurations above the Curie temperature, while in the itinerant case, the moment size is strongly dependent on the specific pattern, and generally the moments are strongly suppressed above the ordering temperature. From an experimental point of view, the susceptibility of itinerant magnetic materials either does not follow Curie-Weiss behavior over a substantial temperature range above the ordering temperature or else it can be fit by Curie-Weiss, but only with a moment size that is inconsistent with the ordered moment.

**Table 1** | Calculated magnetic energies $E_{mag}$ for CrSi$_2$ and spin magnetizations, M FeSi$_2$ at selected virtual crystal doping levels, x, on a per formula unit basis. Here positive x is n-type doping and negative x is p-type and $E_{mag}$ is defined as the energy difference between a constrained non-spin-polarized state and the ferromagnetic state

| | x (f.u.$^{-1}$) | M ($\mu_B$) | $E_{mag}$ (meV/f.u.) |
|---|---|---|---|
| CrSi$_2$ | −0.4 | 0.000 | 0.0 |
| CrSi$_2$ | −0.2 | 0.117 | 0.5 |
| CrSi$_2$ | 0.2 | 0.200 | 7.1 |
| CrSi$_2$ | 0.4 | 0.400 | 23.1 |
| FeSi$_2$ | −0.4 | 0.295 | 6.2 |
| FeSi$_2$ | −0.2 | 0.133 | 1.1 |
| FeSi$_2$ | 0.2 | 0.200 | 6.7 |
| FeSi$_2$ | 0.4 | 0.400 | 21.3 |

In actual materials there is a spectrum of behavior between the limits of local moment and itinerant magnetism and there are many materials, such as bcc Fe that show features of both itinerant and local moment magnetism[36]. Based on the fact that no moments are stabilized for antiferro-magnetic order, we conclude the the present case is closer to the itinerant limit. This is consistent with the expected behavior of low carrier density metals.

Finally, we briefly comment on the relationship of magnetism and thermoelectric properties. As mentioned, these materials are expected to have good performance at high temperature, where the magnetic order will be lost. For samples that are in the doping range where ferromagnetism occurs, there will be a magnetic transition as one cools. In that case, additional spin fluctuation scattering is expected near the transition, which will harm the thermoelectric performance in that range. Below the ordering temperature, the DOS will be spin-polarized. The result of this is generically a reduction of the thermopower as was discussed in detail for $Na_xCoO_2$[37]. Interestingly, near the transition, which is often second order or near second order in Stoner systems, one may have a high paramagnetic susceptibility. In that case, one may expect a magnetothermopower effect in which applied field can significantly reduce the thermopower.

To summarize, we find itinerant ferromagnetism in doped $β$-$FeSi_2$ and $CrSi_2$. The magnetism is half-metallic with an onset at relatively modest doping levels for $n$-type, especially for $β$-$FeSi_2$, and is weaker for $p$-type. This magnetism is connected with the heavy mass, anisotropic bands at the band edges, and the high transition element $d$ character of these bands. It is interesting to note that these features are also connected with the thermoelectric properties of the materials, similar to $Na_xCoO_2$[38–41]. It will be of interest to determine whether the doping levels suggested here can be achieved, and if so whether ferromagnetic quantum critical points and a crossover to ferromagnetic behavior similar to that in $FeGa_3$ can be found in $β$-$FeSi_2$ or $CrSi_2$.

## Methods

We did first principles calculations with the generalized gradient approximation of Perdew, Burke and Ernzerhof (PBE)[42], and the general potential linearized augmented planewave (LAPW) method[43], as implemented in the WIEN2k code[44]. Well converged basis sets, including local orbitals to treat the transition element semicore states were employed along with dense Brillouin zone samplings. These were a minimum of 24 × 24 × 16 for hexagonal $CrSi_2$ and 16 × 16 × 14 for orthorhombic $β$-$FeSi_2$. LAPW sphere radii of 2.5 Bohr and 2.0 Bohr were used for Cr and Si, respectively in $CrSi_2$. For $β$-$FeSi_2$, we used sphere radii of 2.32 Bohr and 2.06 Bohr for Fe and Si, respectively. Relativity was treated at the scalar relativistic level. In both compounds we used the experimental lattice parameters and relaxed the internal coordinates of the atoms by total energy minimization.

$CrSi_2$ is hexagonal, space group $P6_222$ (180), $a$ = 4.4276 Å and $c$ = 6.3681 Å[17]. The Cr is on symmetry site 3c, (1/2,0,0), and the relaxed positions of the Si (6i) are (0.1664,0.3328,0.000), which is very close to the reported experimental position of (0.1656,0.3312,0.000)[17]. $β$-$FeSi_2$ is orthorhombic, space group $Cmca$ (64), $a$ = 9.863 Å, $b$ = 7.833 Å and $c$ = 7.791 Å[45]. The relaxed atomic positions are: Fe1(8d) at (0.2163,0.5,0.5), Fe2(8f) at (0.0,0.6920,0.3119), Si1(16 g) at (0.1286,0.2265,0.4496) and Si2(16 g) at (0.3737,0.4550,0.2739), also in close accord with the reported experimental values.

Magnetism was investigated using virtual crystal approximation (VCA) calculations. In the VCA calculations doping was done by changing the electron count along with the effective atomic number on the Si site. Specifically, the electron count was increased or decreased, for $n$-type or $p$-type doping, respectively. This is an average potential approximation and allows for continuous variation of the doping level. We also did calculations for $β$-$FeSi_2$, where one of the Si atoms in the 24 atom cell was replaced by B ($p$-type), P ($n$-type) or As ($n$-type). In this case the atomic coordinates were again relaxed by total energy minimization. This yielded similar results to the virtual crystal for the corresponding doping level.

## Acknowledgments

Work at ORNL was supported by the Department of Energy, Basic Energy Sciences, Materials Sciences and Engineering Division.

## Author contributions

D.J.S. did first principles calculations for the ferromagnetic cases. D.P. checked for antiferromagnetism. Both authors contributed to the concept and participated in drafting the manuscript.

## Additional information

**Competing financial interests:** The authors declare no competing financial interests.

**How to cite this article:** Singh, D.J. & Parker, D. Itinerant magnetism in doped semiconducting $\beta$-$FeSi_2$ and $CrSi_2$. *Sci. Rep.* **3**, 3517; DOI:10.1038/srep03517 (2013).